\pdfoutput=1

\documentclass[pra,10pt,superscriptaddress,footnoteinbib]{revtex4}
\usepackage{amsmath}
\usepackage{latexsym}
\usepackage{amssymb}
\usepackage{graphics,epstopdf}
\usepackage[colorlinks=true, citecolor=blue, urlcolor=blue ]{hyperref}
\usepackage{epsf,graphics,graphicx}

\newcommand{\n}{\noindent}

\newcommand{\ed}{\end{document}}
\newcommand{\beq}{\begin{equation}}
\newcommand{\eeq}{\end{equation}}

\begin{document}
\title{Effect of cosmic string on spin dynamics}
\author{Debashree Chowdhury\footnote{Electronic address:{debashreephys@gmail.com}}${}^{}$  and B. Basu\footnote{Electronic
address: {sribbasu@gmail.com}} ${}^{}$} \affiliation{Physics and
Applied Mathematics Unit, Indian Statistical Institute, 203
B.T.Road, Kolkata 700 108, India}


\begin{abstract}
\n
In the present paper, we have investigated the role of cosmic string on spin current and Hall electric field. Due to the background cosmic string, the modified electric field of the system generates renormalized spin orbit
coupling, which induces a modified non-Abelian gauge field. The defect causes a change in the AB
and AC phases appearing due to the modified electromagnetic field. In addition, for a time varying
electric field we perform explicit analytic calculations to derive the exact form of spin electric field
and spin current, which is defect parameter dependent and of oscillating type. Furthermore, in an
asymmetric crystal within the Drude model approach we investigate the dependence of the cosmic
string parameters on cosmic string induced Hall electric field.
\end{abstract}


\maketitle
\section{Introduction}
The role of topological defect on physical properties of electron is the subject of study in many areas of physics from condensed matter to high energy.
In classical filed theory, topological defect appears as a consequence of spontaneously
broken symmetry. Domain walls, strings, monopoles and textures are various types of topological defects among which cosmic strings are of special interest. Cosmic strings are important ingredients for the generation of enormous interesting physical phenomena and establish the link between condensed matter and high energy physics, which certainly is one of the promising topic of attraction  \cite{ bakke, Bakke1, bakke2, furtado,marques}. The effect of rotation on Aharonov-Bohm (AB) and Aharonov-Casher(AC) phases in presence of topological defect are studied in \cite{bakke,Bakke1,bakke2}. In \cite{furtado,marques}, the authors have discussed the effect of topological defect on Landau levels and how the infinite degeneracy of the Landau levels breaks in presence of the cosmic string is also shown. In this respect, it is of interest to study the spin transport phenomenon in semiconductors in the presence of cosmic string background.

The cosmic defect can be regarded as the bridge between the high energy physics and spin dynamics. Recently the
study of spin has open up a new branch of physics, spintronics.
Semiconductor spintronics \cite{wolf,zutic,sh1} 
is one of the hottest area of research now a days. In semiconductor spintronics, the spin orbit coupling ($SOC$) plays a crucial role. $SOC$, which originates from the relativistic coupling of spin and orbital motion of electrons is a very important concept for the study of all spin related issues in semiconductor. The induced SOC can be achieved by various means and has attracted similar attention \cite{bc, cb, bcs, matsuo} as the usual SOC term in semiconductors. The achievement of $SOC$ through the spin rotation coupling is discussed in a recent paper \cite{sr}.
In $spintronics$, the challenging issue is the detection and control of spin current .
It was Dyakonov and Perel \cite{perel}, who pointed out that similar to the classical Hall effect by electron charge,
electron spins also show the spin Hall effect($SHE$). SHE is described in different occasion in different systems. Recently SHE is also discussed in the tilted electron vortex beams \cite{pro}. $SHE$ is the mechanism to achieve spin polarized current. We have encountered the problem through gauge theoretical approach in a cosmic string space time background. One of our main motivations is to study the spin current in presence of cosmic defect through gauge theoretical approach as the gauge theory is a popular tool in deriving the spin Hall current.

A simple description of $SHE$ can be achieved by the Drude model approach \cite{n}. In \cite{n} it has been shown that for a special symmetry of the crystal how the spin current and conductivity depend on the symmetry parameters. In \cite{cb,bc}, we have studied the effect of acceleration on spin current and on Berry curvature in semiconductor in the Drude model approach taking into account the cubic symmetry. However, in real experiments, one have to deal with the non-cubic crystals. There exists a dependence of crystal symmetry and geometry of experiments on intrinsic spin Hall effect \cite{noncubic}. This motivates us to investigate the effect of crystal symmetry and  geometry on spin current in presence of a cosmic string background. In particular, observation of spin current in the presence of topological defect in an asymmetric crystal, which is still overlooked, may provide some impressive results. Furthermore, investigation of how the spin current as well as Hall voltage change with the parameters of cosmic string is also an interesting scenario to investigate.

Before going further, let us briefly mention about the orientation of the paper. In
section II, we discuss the Hamiltonian of a Dirac electron in the cosmic string background. Section III deals with the corresponding AB and AC phases in the system. In section IV, we present the concept of spin dependent electric field as well as the spin current for an external time dependent electric field in the presence of cosmological string background. How the  Hall electric field in an asymmetric crystal is modified due to the presence of topological defect is narrated in section V. Finally, we conclude in section VI.

\section{Effective SOC Hamiltonian in presence of cosmic string}

We first build our model Hamiltonian and then proceed to find the conditions under which we get the spin current for our system. To write the Dirac Hamiltonian in presence of cosmic string background we have to take into account the minimal coupling term due to the interaction between the charged spinor and electromagnetic fields involved in the system. This in turn induces a non trivial metric. We can write the covariant Dirac equation in curved space time as \cite{naka}
\beq [\tilde{\gamma}^{\mu}(x)(p_{\mu} - qA_{\mu}(x) - \Gamma_{\mu}(x)) + mc^{2}]\phi(x) = 0,\label{a}\eeq
where $A_{\mu}$ is the electromagnetic gauge potential, $\Gamma_{\mu}(x)$ is the spin connection and the $\tilde{\gamma}^{\mu}$ form a Clifford algebra in the curved space time as $\{\tilde{\gamma}^{\mu}, \tilde{\gamma}^{\mu}\} = 2g^{\mu\nu}(x).$
Here, $g^{\mu\nu}(x)$ is the metric tensor of space time in presence of the background topological defect. 
We can write the line element as
\beq ds^{2} = c^{2}dt^{2} - d\rho^{2} - \eta^{2}\rho^{2}d\phi^{2} - dz^{2}, \eeq
where $\eta = 1 - \frac{4\lambda G}{c^{2}}$ and $\lambda$ are the deficit angle and linear mass density of the cosmic string respectively and $\rho,$ $\phi,$ and z are standard cylindrical polar coordinates and t is the time component. Usually the deficit angle takes values $\eta> 1$ which corresponds to the anti-conical space time having negative curvature. In the tetrad formalism the metric tensor can be written as
\beq g_{\mu\nu}(x) = e^{m}_{~~\mu}(x)e^{n}_{~~\nu}(x)\eta_{mn},\eeq where $e_{~~m}^{\mu}$'s are the vierbeins which obey the relation $e^{m}_{~~\mu}(x)e^{\mu}_{~~n}(x) = \delta^{m}_{~~n}$ \cite{bakke} and $\eta_{mn}$ is the metric of the flat space. The vierbeins can be represented as
\begin{eqnarray}\label{om}
e^{\mu}_{~~m}  = \left(\begin{array}{ccrr}
1 & 0 & 0 & 0 \\
0 & cos\varphi & sin\varphi & 0\\
0 &-\frac{sin\varphi}{\eta\rho}& \frac{cos\varphi}{\eta\rho} & 0 \\
0 & 0 & 0 & 1
\end{array} \right),~~~~~
e_{~~\mu}^{m}  = \left(\begin{array}{ccrr}
1 & 0 & 0 & 0 \\
0 & cos\varphi & -\eta\rho sin\varphi & 0\\
0 & sin\varphi & \eta\rho cos\varphi & 0 \\
0 & 0 & 0 & 1
\end{array} \right).
\end{eqnarray}
We can get back the flat space time for $\eta = 1.$
The spinor connection are connected with the vierbeins as
\beq \Gamma_{\mu} = \frac{1}{8}e_{m\nu}D_{\mu}e^{\nu}_{~~n}[\gamma^{m}, \gamma^{n}], \eeq where $D_{\mu} = \partial_{\mu} + \Gamma_{\mu},$ is the covariant derivative due to the background space time. In our case the only nonzero component is $\Gamma_{\phi}$ as all other components vanishes \cite{bakke}. Thus we can write 
\beq \Gamma_{\phi} = -\frac{i}{2}(1 - \eta)\Sigma_{3},\eeq where $\Sigma_{3}$ is given by
\begin{equation}\label{dd}
\Sigma_{3} = \left(\begin{array}{cr}
\sigma_{3} & 0  \\
0 & \sigma_{3}
\end{array} \right),
\end{equation}
with $\sigma_{3}$ as the usual Pauli matrix.
From (\ref{a}), we can write the deformed Dirac equation of an electron in the cosmic string background as \cite{ma}
\begin{eqnarray}\label{b}
H = \beta mc^{2} +  c\left(\vec{\alpha}.\vec{\pi}\right) +
qV(\vec{r}) + \vec{\alpha}.\vec{\Gamma} + c\left(\vec{\alpha}.\vec{\vec{\Omega}}.\vec{\pi}\right) + \Gamma_{0},
\end{eqnarray}
where $\vec{\pi} = \vec{p} - \frac{q\vec{A}}{c}$ is the minimal momentum of the particle in presence of the external magnetic field $\vec{B} =\vec{\nabla}\times\vec{A}.$ $\vec{\alpha}, \beta$ are the Dirac matrices and for convenience one can define  $\vec{\vec{\Omega}}$ as $\Omega^{m}_{~~\mu}(x) = e^{m}_{~~\mu}(x) - \delta^{m}_{~~\mu}$ and $\Omega^{\mu}_{~~m}(x) = e^{\mu}_{~~m}(x) - \delta^{\mu}_{~~m},$ where $\delta^{\mu}_{~~m}$ represents the Minkowski frame in the cylindrical polar coordinate.
 First term in the r.h.s  of (\ref{b}), is the rest energy term, 2nd term is kinetic energy term, third term is due to external electric potential. The next two terms $\vec{\alpha}.\vec{\Gamma}$ and  $\left(\vec{\alpha}.\vec{\vec{\Omega}}.\vec{\pi}\right)$ are the hidden momentum term and geometry related term of the cosmic string respectively. Here in (\ref{b}), we have neglected the second order terms of $\left(\vec{\alpha}.\vec{\vec{\Omega}}.\vec{\pi}\right).$ The last term $\Gamma_{0}$, is an electric potential like term and does not contribute anything in our system and will be dropped for future calculations. One can easily conclude from (\ref{b}) that the momentum is modified not only due to the external magnetic field but also due to the presence of cosmic string in the background. 

Eqn. (8) can now be rewritten in terms of the total modified momentum as
\begin{eqnarray}\label{bc}
H =\beta mc^{2} +  c\vec{\alpha}.\left(\vec{p} - \frac{q}{c}\vec{A}_{tot}\right) + qV(\vec{r}),
\end{eqnarray}
where $\vec{A}_{tot} = \left(\vec{A} +\frac{\vec{\Gamma}}{q} +\frac{c\vec{\vec{\Omega}}.\vec{\pi}}{q}\right) = \left(\vec{A} + \vec{A}_{cs}\right),$ is the total gauge field in the system with $\vec{A}_{cs} = \left(\frac{\vec{\Gamma}}{q} +\frac{c\vec{\vec{\Omega}}.\vec{\pi}}{q}\right)$ as the gauge due to the cosmic string background.\\

The effect of the cosmic string on the spin dynamics can be well observed from the Dirac Hamiltonian in the non-relativistic limit where the spin-orbit interaction term comes into play explicitly. So to proceed further we express the Hamiltonian of the Dirac particle in the non relativistic limit
 using the block diagonalization method exploiting FWT \cite{m}.
 The Hamiltonian (\ref{b}) can be divided into  block diagonal and off diagonal parts respectively denoted by $\epsilon $ and $ O $ as
 \begin{eqnarray}\label{c}
 H &=& \beta mc^{2} + O +\epsilon ,\nonumber\\
  O &=& c\vec{\alpha}.((\vec{p}-\frac{q\vec{A}}{c})+ \vec{\alpha}.\vec{\Gamma} + c\left(\vec{\alpha}.\vec{\vec{\Omega}}.\vec{\pi}\right),\nonumber\\
 \epsilon &=& qV(\vec{r}).
 \end{eqnarray}
 The non relativistic FW transformed Hamiltonian in terms of $ \epsilon $ and $ O $ is given by
 \begin{equation}
 H_{FW} = \beta \left(mc^{2}+\frac{O^{2}}{2mc^{2}}\right)+ \epsilon
 -\frac{1}{8m^{2}c^{4}} \left[O ,[O,\epsilon]\right]
 \end{equation}
Neglecting the rest energy term the Pauli- Schroedinger equation in presence of cosmic string can be readily obtained as \cite{ma}
\begin{eqnarray}
 H_{FW} &=&  \frac{1}{2m}\left((\vec{p}-\frac{q\vec{A}_{tot}}{c})\right)^{2} + qV(\vec{r}) +  -\frac{q\hbar}{2mc}\vec{\sigma}. \vec{B} - \frac{q\hbar}{2mc}\vec{\sigma}. \left(\vec{\nabla}\times \vec{A}_{cs}\right) + \frac{q\hbar}{4m^{2}c^{2}}\vec{\sigma}.(\vec{\nabla}V(\vec{r})\times \vec{p})\nonumber\\
 && + \frac{q\hbar}{4m^{2}c^{2}}\vec{\sigma}.(\vec{\vec{\Omega}}.\vec{\nabla}V(\vec{r})\times \vec{p}) + \frac{q\hbar^{2}}{8m^{2}c^{2}}\vec{\nabla}.\vec{E} + \frac{q\hbar^{2}}{8m^{2}c^{2}}\vec{\nabla}.\left( -\vec{\vec{\Omega}}.\vec{\nabla}V(\vec{r})\right) \nonumber\\ 
 &=& H_{k} + H_{B} + H_{SO} + H_{d}.
 \end{eqnarray}
All parts of the Hamiltonian are explicitly written and discussed below. \\
 The kinetic part $H_{k}$ is obtained as
\beq H_{k} = \frac{1}{2m}\left((\vec{p}-\frac{q\vec{A}_{tot}}{c})\right)^{2} + qV(\vec{r}).\eeq
In this kinetic part the dependence of the cosmic string parameters are implicit in the first term via $\frac{\Gamma}{c}$ and  $\vec{\vec{\Omega}}.\vec{\pi}$ terms. Explicitly this terms are incorporated in the gauge field term $\vec{A}_{cs}.$ But one can show the dependence explicitly as well by taking into account 
\begin{equation}\label{r}
\Omega^{\mu}_{~~a}  = \left(\begin{array}{ccrr}
0 & 0 & 0 & 0 \\
0 & cos\varphi & sin\varphi & 0\\
0 &\frac{sin\varphi}{\eta\rho}& -\frac{cos\varphi}{\eta\rho} & 0 \\
0 & 0 & 0 & 0
\end{array} \right).
\end{equation} When we consider the term $\frac{\Gamma}{c}, $ the only possible contribution due to cosmic string is due to the term $ \Gamma_{\phi} = -\frac{i}{2}(1 - \eta)\sigma_{3}$ as indicated in equation (6). On the other hand we should note here that $\vec{A}_{cs}$ contains a vector $\vec{L} = \vec{\vec{\Omega}}.\vec{\pi}$ whose matrix representation can be written as

\begin{equation}\label{me}
L_{\mu}  = \left(\begin{array}{ccrr}
0 \\
\pi_{\rho}cos\varphi + \pi_{\varphi}sin\varphi \\
-\frac{\pi_{\rho}}{\eta\rho}sin\varphi+ \frac{\pi_{\varphi}}{\eta\rho}cos\varphi \\
0 
\end{array} \right),
\end{equation}
here $\mu$ denotes different components of the vector $\vec{L}.$ This shows that our kinetic term contains explicitly the effect of cosmic string parameters.
  
The Zeeman like term $H_{B}$ is given by
\beq H_{B} = -\frac{q\hbar}{2mc}\vec{\sigma}. \vec{B} - \frac{q\hbar}{2mc}\vec{\sigma}. \vec{B}_{cs} = -\frac{q\hbar}{2mc}\vec{\sigma}. \vec{B}_{tot}\eeq
where $\vec{B}_{cs} = \vec{\nabla}\times\vec{A}_{cs} $ is the effective magnetic field generated due to the cosmic string background and $ \vec{B}_{tot} = \vec{B} + \vec{B}_{cs},$ is the total magnetic field effective in the system. Here also we can express the dependence of cosmic string on $\vec{B}_{cs}$ as 
\begin{equation}
\vec{B}_{cs} = \frac{c}{q\rho}\left[\pi_{\rho}\left(cos\phi+sin\phi-\frac{sin\phi}{\eta\rho}\right) + \pi_{\rho}\left(sin\phi+\frac{cos\phi}{\eta\rho}- cos\phi\right) \right]\hat{z}.
\end{equation}
This definitely suggests that the field $\vec{B}_{cs}$ acts along the $z$ direction. Also the dependence of the defect parameters are clear from the expression in (18). 

It may be noted that similar to the effective magnetic field, the electric field also gets modified due to the topological defect present in the background. The defect induced electric field generates a SOC term and modifies the SOC term due to the external electric field. As the SOC term is a crucial parameter for analyzing all spin related issues in spintronic applications, it is relevant to write the terms within total effective spin-orbit Hamiltonian explicitly. The modified effective SOC term with the cosmic string background is expressed as
\begin{eqnarray}
H_{SO} &=& \frac{q\hbar}{4m^{2}c^{2}}\vec{\sigma}.(\vec{\nabla}V(\vec{r})\times \vec{p})
 + \frac{q\hbar}{4m^{2}c^{2}}\vec{\sigma}.(\vec{\vec{\Omega}}.\vec{\nabla}V(\vec{r})\times \vec{p}) \nonumber\\
 &=& -\frac{q\hbar}{4m^{2}c^{2}}\vec{\sigma}.\left(\left(\vec{E} + \vec{E}_{cs}\right)\times \vec{p}\right)\nonumber\\
 &=& -\frac{q\hbar}{4m^{2}c^{2}}\vec{\sigma}.\left(\vec{E}_{tot}\times \vec{p}\right)
\end{eqnarray}
where $\vec{E}_{cs} =  -\vec{\vec{\Omega}}.\vec{\nabla}V(\vec{r}),$ and  $\vec{E}_{tot} = (\vec{E} + \vec{E}_{cs}),$ is the total electric field in the system. One can write the induced electric field $\vec{E}_{cs}$ as
\begin{equation}
E_{cs,\mu}= \left(\begin{array}{ccrr}
0\\
\vec{\nabla}_{\rho}V(\vec{r})cos\varphi + \vec{\nabla}_{\varphi}V(\vec{r})sin\varphi\\
-\frac{\vec{\nabla}_{\rho}V(\vec{r})}{\eta\rho}sin\varphi+ \frac{\vec{\nabla}_{\varphi}V(\vec{r})}{\eta\rho}cos\varphi
\\
0 
\end{array} \right),
\end{equation}
 It is evident from the expression of the cosmic string induced electric field $\vec{E}_{cs}$ that it depends on the external potential as well as the cosmic string parameters $\eta$ also. Though the defect induced electric field $\vec{E}_{cs}$ depends on the defect parameter $\eta$, still it can be somewhat controlled via the applied external field. How this renormalized SOC affects the spin current, spin electric field and Berry curvature is the topic of our interest and is elaborated in the next sections.

Lastly, in (13), the Darwin term is as follows
\beq H_{d} = \frac{q\hbar^{2}}{8m^{2}c^{2}}\vec{\nabla}.\vec{E} + \frac{q\hbar^{2}}{8m^{2}c^{2}}\vec{\nabla}.\vec{E}_{cs} = \frac{q\hbar^{2}}{8m^{2}c^{2}}\vec{\nabla}.\vec{E}_{tot}  ,\eeq
which does not contribute for a constant electric field.\\

The relevant part of the Hamiltonian (12) can now be written as
\begin{eqnarray}\label{e}
H =  \frac{(\vec{p} - \frac{q\vec{A}_{tot}}{c})^{2}}{2m} + qV(\vec{r}) - \frac{q\hbar}{2mc}\vec{\sigma}. \vec{B}_{tot}
- \frac{q\hbar}{4m^{2}c^{2}}\vec{\sigma}. \left(\vec{E}_{tot}\times \vec{p}\right), 
\end{eqnarray}
where $\vec{A}_{tot} = \vec{A} + \vec{A}_{cs},$ $B_{tot} = \vec{\nabla}\times \vec{A}_{tot}$ and $\vec{E}_{tot} =  (\vec{E} -\vec{\vec{\Omega}}.\vec{\nabla}V(\vec{r}),$ are the total gauge field , total magnetic and total electric fields of the system respectively.
The Hamiltonian in (\ref{e}) is our model Hamiltonian which we use in the following sections to derive many interesting spin related quantities. This Hamiltonian in (\ref{e}) have $SU(2)\times U(1)$ symmetry, where the $SU(2)$ gauge is given by $a^{\mu} = (a^{0}, \vec{a}) = (-\vec{\sigma} . \vec{B}_{tot}, \vec{\sigma}\times \frac{\vec{E}_{tot}}{2} ),$  and the U(1) gauge potential is $A^{\mu} = (V_{tot}, \vec{A}_{tot}).$
The U(1) gauge field is associated with the magnetic field due to externally applied one as well as the magnetic field appearing as a consequence of cosmic string background. The vector potential in the U(1) gauge is not only a mathematical instrument to describe the magnetic field, rather it can manifest itself as a phase, known as Aharonov-Bohm $(AB)$ phase \cite{aharonov}. In the next section we will discuss about the modification in the AB and AC phases respectively due to the presence of the cosmic string. 
\section{Modified AB and AC phases}
In this section we first concentrate on the AB phase of the system. The AB phase can be calculated using Hamiltonian (\ref{bc}) or Hamiltonian (21). This is so because of the fact that FW transformation does not change the effective gauges of the system. Whether we start from (\ref{bc}) or (21) we can write the Aharonov Bohm phase (AB) as follows
\beq \phi^{cs}_{AB} = \oint_{c} d\vec{r}.\vec{A}_{tot}(r),\eeq
which contain the effect of both external magnetic field and the background cosmic string. In our system the corresponding $AB$ phase, which may be termed as defect modified AB phase can be written as
\beq \phi^{cs}_{AB} = \oint_{c} d\vec{r}.\vec{ A}_{tot}(r) =  \oint_{c} d\vec{r}.\vec{A}(r) +  \oint_{c} d\vec{r}.\vec{A}^{cs}(r) \label{ab},\eeq where $\vec{A}^{cs}(r) = \frac{\vec{\Gamma}}{q} + \frac{c}{q}\vec{\vec{\Omega}} . \vec{\pi}.$ It is clear from (\ref{ab}), that the effect of topological defect is incorporated in the $AB$ phase through the second term in (23) via the gauge potential $\vec{ A}^{cs}(r).$ Thus we have the modification of the AB phase due to the cosmic string effects as

\beq \phi^{mod,~cs}_{AB} =  \oint_{c} d\vec{r}.\vec{A}^{cs}(r) \label{abv}.\eeq \\

Next to attain the AC phase, we proceed in the following manner.
In cosmic string background using the Hamiltonian in (\ref{e}), we can calculate the force equation for the carriers with cosmic string background. Let us start with the evaluation of the velocity. Following the Heisenberg equation we can write the expression of velocity as
\beq \vec{\dot{r}}  = \frac{1}{i\hbar}[\vec{r}, H ] = \frac{\vec{p}}{m} -
\frac{q\hbar}{4m^{2}c^{2}}\left(\vec{\sigma}\times \vec{E}_{tot}\right)\label{deb}.\eeq Similarly,
\beq \vec{\dot{p}} = \frac{1}{i\hbar}[\vec{p}, H]= -q\nabla V_{tot} +
\frac{q\hbar}{4m^{2}c^{2}}\nabla\left[(\vec{\sigma}\times \vec{E}_{tot}).\vec{p}\right].\eeq From equation (\ref{deb}), one can write
\beq \vec{p} = m \vec{\dot{r}} - \frac{q\hbar}{4mc^{2}}\sigma\times \vec{E}_{tot}. \label{dea}\eeq
From (\ref{deb}) and (\ref{dea}), we can write the force equation as
\beq \vec{F} = m\ddot{\vec{r}} = -q\nabla V_{tot} + \frac{q\hbar}{4mc^{2}}\vec{\dot{r}}\times\left[\nabla\times\sigma\times\vec{E}_{tot}\right] = \vec{F}_{0} + \vec{F}_{\sigma}.\eeq This eqn. is analogous to the Lorentz force equation in classical Hall effect. The second term in the force equation is the spin dependent part, responsible for the separation of spin.
One can write the total Lorentz force equation as
\beq \vec{F}_0 + \vec{F}_{\vec{\sigma}} = q\vec{E}_{tot} + \frac{q}{c}(\dot{\vec{r}} \times\vec{B}_{tot}^{cs}(\vec{\sigma}))\label{sigma},\eeq where $\vec{B}_{tot}^{cs}(\sigma) = \left[\nabla\times\sigma\times\vec{E}_{tot}\right],$ is the magnetic field appearing in the rest frame of the electron due to the SOC term. Here one can write the corresponding spin Lorentz force is
\beq  \vec{F}_{\vec{\sigma}} =  \frac{q}{c}(\dot{\vec{r}} \times \vec{B}_{tot}^{cs}(\vec{\sigma}))\label{sigma1}.\eeq
This spin dependent Lorentz force is responsible for the spin current produced in the system. $ \vec{B}_{tot}^{cs}(\vec{\sigma})$, the $effective$ $magnetic$ $field$ appearing in the spin space can be read as
\begin{eqnarray}\label{asigma}
\vec{B}_{tot}^{cs}(\vec{\sigma}) = \vec{\nabla}\times {\cal\vec{A}}^{cs}(\vec{\sigma})
\end{eqnarray}
where the vector potential is explicitly given by
\begin{eqnarray}
{\cal\vec{A}}_{tot}^{cs}(\vec{\sigma}) =   -\frac{\hbar}{4mc}(\vec{\sigma}\times \vec{E}_{tot}).
\end{eqnarray}
It can be noted that the gauges $\vec{A}_{cs}$ and ${\cal\vec{A}}_{tot}^{cs}$ are not the same gauge field. The first one is U(1) gauge arising in the system due to the cosmic string background. The latter is a SU(2) gauge potential appears due to the total SOC term which arises from the external electric field as well as cosmic string induced electric field $\vec{E}_{cs}.$ 

The Hamiltonian in (21) can be written in terms of this spin dependent gauge for the total electric field $\vec{E}_{tot}$
\begin{eqnarray}\label{kj}
H =  \frac{(\vec{p} - \frac{q\vec{A}_{tot}}{c} - \frac{q}{c} {\cal\vec{A}}_{tot}^{cs}(\sigma))^{2}}{2m}.
\end{eqnarray}
Here, we have neglected the second order terms of $ {\cal\vec{A}}_{tot}^{cs}.$

The AB phase in (23), which contains the effect of cosmic string  is same for both up and down electrons, whereas AC phase \cite{casher}, which can be written as  \beq \phi^{cs}_{AC} = \oint d\vec{r}. {\cal\vec{A}}_{tot}^{cs}( \sigma),\eeq is equal but opposite for the two different spin orientations.

As the AC phase contains ${\cal\vec{A}}_{tot}^{cs}( \sigma)$ term which contains the defect parameter through $\vec{E}_{cs},$ we can rename this AC phase as defect modified AC phase and can be written as
\beq \phi^{cs}_{AC} = \oint d\vec{r}. {\cal\vec{A}}_{tot}^{cs}( \sigma) =  \oint d\vec{r}. {\cal\vec{A}}_{o}(\sigma) +  \oint d\vec{r}. {\cal\vec{A}}_{cs}(\sigma),\eeq where ${\cal\vec{A}}_{o}(\sigma) = -\frac{\hbar}{4mc}(\vec{\sigma}\times \vec{E})$ with $\vec{E}$ as the external electric field and ${\cal\vec{A}}_{cs}(\sigma) = -\frac{\hbar}{4mc}(\vec{\sigma}\times \vec{E}_{cs})$ with $\vec{E}_{cs}$ being the electric field due to the cosmic string background. Thus we have the modification of the AC phase due to the cosmic string effects as
\beq \phi^{mod,~cs}_{AC} = \oint d\vec{r}. {\cal\vec{A}}_{cs}(\sigma) = -\frac{\hbar}{4mc}\oint d\vec{r}. (\vec{\sigma}\times \vec{E}_{cs})\eeq 
Thus we can argue that both AB and AC phases are modified due to the presence of cosmic defect. From eqn. (23) and (35) we can demonstrate that both AB and AC phases are modified due to the presence of cosmic defect. Furthermore, following \cite{bc}, through the interplay of the AB and AC phases, both of which contain the effect of cosmic string, one can theoretically propose and explain the configuration of spin filter \cite{hatano}.

\section{Defect induced alternating spin current}
Considering only $SU(2)$ gauge symmetry with the Zeeman term and with the help of eqn. (32), the total Hamiltonian in (21) can be written as
 \begin{eqnarray}\label{km}
H =  \frac{(\vec{p} - \frac{q}{c}{\cal \vec{A}}^{cs}(\sigma))^{2}}{2m} - \frac{q\hbar}{2mc}\vec{\sigma}. \vec{B}_{tot}.
\end{eqnarray}
In terms of the $SU(2)$ gauge  $a^{\mu} = (a^{0}, \vec{a}) = (-\vec{\sigma} . \vec{B}_{tot}, {\cal\vec{A}}^{cs}(\sigma)  ),$  the curvature term is given by
\beq {\displaystyle F}_{\mu\nu} = \partial_{\mu}a_{\nu} - \partial_{\nu}a_{\mu} - \frac{ie}{\hbar}[a_{\mu},a_{\nu}],\eeq
where $\mu, \nu = {t, x, y, z}$ are the space time coordinates. As the SU(2) vector potential $a^{\mu}$  is non-Abelian in nature,
so to find out the curvature term, we have to consider the non commutative contribution of different components of vector potentials.
The spatial part of the field tensor gives the Berry curvature\cite{berry},  which effectively corresponds to the magnetic field whereas the temporal part can give rise to the spin electric field.

Our next job is to find out the spin dependent electric field from the temporal part of the field tensor. The electric part of the field tensor can be obtained as
\beq {\cal E}_{i} = {\displaystyle F}_{i0} = \partial_{i}a_{0} - \partial_{0}a_{i} - \frac{iq}{\hbar}[a_{i}, a_{0}].\label{ss}\eeq
Using equations (22,32,37), we can define a new parameter $\alpha_{tot}$ as $\alpha_{tot} = \frac{\hbar}{4m^2c^{2}}E_{tot}^{z},$ where the total electric field is considered along the $z$ direction. This is very similar to the well known Rashba SOC in semiconductor Heterostucture \cite{cb, bc}. 
The $i ^{th}$ component of this electric field, considering only the $z$ component of the magnetic field is as follows
\beq {\cal E}_{i} =  -\frac{\partial B_{tot,z}}{\partial x_{i}}\sigma^{z} + \frac{mc}{\hbar}\frac{\partial \alpha_{tot}}{\partial t}\sigma^{j}
- \frac{imqc}{\hbar^{2}}\alpha_{tot}B^{z}_{tot}\left[\sigma^{j},\sigma^{z}\right],\label{ca}\eeq where $i, j = \rho, \phi.$ 
Using the commutation relation between the Pauli matrices we can write 
\beq {\cal E}_{i} =  -\frac{\partial B_{tot,z}}{\partial x}\sigma^{z} + \frac{mc}{\hbar}\frac{\partial \alpha_{tot}}{\partial t}\sigma^{j} + \frac{2mqc}{\hbar^{2}}\alpha_{tot}B^{z}_{tot}\sigma^{i},\label{va}\eeq
$\vec{B}_{tot},$ the effective magnetic field being is constant in magnitude we
 can neglect the first term in ${\cal E}_{i}.$
The spin dependent part of the $i$th component of electric field as
\beq {\cal E}_{i}^{\sigma} = -\frac{imc}{\hbar}\frac{\partial \alpha_{tot}}{\partial t}\sigma^{z} = {\cal E}\sigma^{z}\label{vab},\eeq
where ${\cal E} = -\frac{imc}{\hbar}\frac{\partial \alpha_{tot}}{\partial t}.$ In a simplified form we can write \beq {\cal E}^{\sigma}_{i} = \mp\frac{imc}{\hbar}\frac{\partial \alpha_{tot}}{\partial t}.\eeq
Here $\mp$ sign corresponds to the electron having spins parallel and antiparallel to $z$ axis. The effective spin dependent electric field ${\cal E}^{\sigma}_{i}$ results in a spin
separation along $z$ direction and generates a spin current.
Obviously, the spin dependent part of the electric field i.e the Yang mills electric field is proportional to the time derivative of the Rashba like SO coupling $\alpha_{tot},$ which effectively contains the total electric field $\vec{E}_{tot}$ One may note that the motion of spins can be controlled by controlling the effective time dependent electric field.
The spin independent part of this electric field depends on the Rashba like coupling parameter $\alpha_{tot}$ and on the total magnetic field is given by
\beq {\cal E}^{c}_{i} = \frac{2mqc}{\hbar^{2}}\alpha_{tot}B^{z}_{tot}.\eeq

Let us now
consider the external electric field as time dependent. As an example of time dependent electric field we consider \beq \vec{E} = E_{0} exp(i\omega t)\hat{n} ,\label{atime}\eeq where the electric field is induced by harmonic oscillation with frequency $\omega$ and amplitude $E_{0}$ and $\hat{n}$ is the unit vector. This special form of the electric field effectively gives the total electric field as \beq E_{tot}^{z} = E_{0}(1 + \vec{\vec{\Omega}}_{z})exp(i\omega t).\eeq  In this case, the spin independent part is given by
\beq {\cal E}^{c}_{i} = \frac{2mqc}{\hbar^{2}}\alpha_{tot}B^{z}_{tot} =  \frac{q}{2mc}B^{z}_{tot}E_{0}(1 + \vec{\vec{\Omega}}_{z})exp(i\omega t)\label{k7} \eeq whereas the spin dependent part can be written as
\beq {\cal E}^{\sigma}_{i} = \pm\frac{\hbar \omega}{4mc}(1 + \vec{\vec{\Omega}}_{z})exp(i\omega t). \label{k8}\eeq
The eqns (\ref{k7}) and (\ref{k8}) indicate that both the electric fields depend on the cosmic string parameter $\vec{\vec{\Omega}}$ and both are of alternating types.

Obviously, this alternating spin dependent electric field can generate an alternating spin current. The total current of the system can be written as a product of spin conductance and the corresponding emf. as
\beq j^{\sigma}_{i} = \vec{\zeta} . \vec{\cal{E}^{'}} = [\zeta_{0}I +\zeta_{s}(\rho .\sigma)].[{\cal E}^{c}_{i}I + {\cal E}^{\sigma}_{i})],\eeq
where $\vec{\rho}$ is the spin polarization axis and  $\vec{\zeta}_{s}$ is the spin conductance.
Thus assuming the polarization axis along $z$ direction, the total current can be written as
\begin{eqnarray}\label{sc}
j_{i} &=& \left[\zeta_{0}I + \zeta_{s}\sigma^{z}].[{\cal E}^{c}_{i}I + ({\cal E})\sigma^{z}\right]\nonumber\\
&=& \left[\frac{1}{2mc}\left(q\zeta_{0}B^{z}_{tot} + i \zeta_{s} \frac{\hbar\omega}{2}\right)I +  \left(q\zeta_{s}B^{z}_{tot} + i \zeta_{0} \frac{\hbar\omega}{2}\right)\sigma^{z}\right]E_{0}(1 + \vec{\vec{\Omega}}_{z})exp(i\omega t)\nonumber\\
&=& j_{i}^{c} + j^{s}_{i}
\end{eqnarray}
where
\begin{eqnarray}
j_{i}^{c} = \frac{1}{2mc}\left(q\zeta_{0}B^{z}_{tot} + i \zeta_{s} \frac{\hbar\omega}{2}\right)E_{0}(1 + \vec{\vec{\Omega}}_{z})exp(i\omega t)I
\end{eqnarray}
is the charge current and
\begin{eqnarray}
j^{s}_{i} = \frac{1}{2mc}\left(q\zeta_{s}B^{z}_{tot} + i \zeta_{0} \frac{\hbar\omega}{2}\right)exp(i\omega t)E_{0}(1 + \vec{\vec{\Omega}}_{z})\sigma_{z}
\end{eqnarray}
is the spin current of the system in the $i^{th}$ direction and both types of current depend on the cosmic string parameter $\vec{\vec{\Omega}}$ and $z$ component of the total magnetic field. But we should mention here that we do not have any contribution from the $z$ component of $\vec{\vec{\Omega}}.$ So what makes the result interesting is the $z$ component of the total magnetic field. As we have calculated in eqn. (17) that the magnetic field explicitly depends on the cosmic parameter $\eta,$ our spin current also explicitly depends on the cosmic parameters. Thus we can explicitly visualize the effect of cosmic string on spin current which is one of the interesting result to notice at. If we do not have the external magnetic field, we still have a finite spin current given by
\begin{eqnarray}
j^{s}_{i} = \pm\frac{E_{0}}{2mc}\left(q\zeta_{s}\frac{c}{q\rho}\left[\pi_{\rho}\left(cos\phi+sin\phi-\frac{sin\phi}{\eta\rho}\right) + \pi_{\rho}\left(sin\phi+\frac{cos\phi}{\eta\rho}- cos\phi\right) \right] + i \zeta_{0} \frac{\hbar\omega}{2}\right)exp(i\omega t),
\end{eqnarray}
where the explicit dependence of cosmic parameters are very clear and $\pm$ denotes the spin up and down contribution respectively.
 
 Furthermore, one can note that the charge and spin currents are not constant but vary with time i.e they are giving the alternating currents in our system. The spin dependent electric field generates as a consequence of the total SOC term and can induce
not only a spin current, but also a charge current in the system.  Observation of this interesting effect \cite{shibata} may be termed as defect induced inverse spin Hall effect.

\section{Defect modified Hall electric field in non-cubic crystal}

One can write Hamiltonian (22) with dynamical terms only as
\beq H = \frac{\vec{p}^{2}}{2m} + qV_{tot}(r) - \frac{q\hbar}{4m^{2}c^{2}}\vec{\sigma}.\left(\vec{E}_{tot}\times\vec{p}\right).\label{kin}\eeq In (\ref{kin}) the last term is the total spin orbit coupling term appearing due to the external electric field and the background cosmic string induced electric field. One can consider that $V(\vec{r}) = V_{0}(\vec{r}) + V_{l}(\vec{r}),$ where $V_{l}(\vec{r})$ is the crystal potential. The Hamiltonian in (\ref{kin}), can then be rewritten as
\beq H = \frac{\left(\vec{p} - \frac{q}{c}{\cal \vec{A}}^{cs}(\sigma)\right)^{2}}{2m} + qV_{tot}(r)\label{21},\eeq where the spin dependent gauge ${\cal \vec{A}}^{cs}(\sigma)$ \cite{fujita, casher} is given by
\beq {\cal \vec{A}}^{cs}(\sigma) = - \frac{\hbar}{4mc}\left(\sigma\times \vec{E}_{tot}\right)\label{q},\eeq where we have neglected the second order terms of ${\cal \vec{A}}^{cs}(\sigma)$ in (\ref{21}). This gauge potential obviously corresponds to a spin dependent magnetic field as \beq \vec{B}^{cs}(\sigma) = - \vec{\nabla}\times {\cal\vec{A}}^{cs}(\sigma) =  \frac{\hbar}{4mc}\left[\vec{\nabla}\times(\vec{\sigma}\times \vec{E}_{tot})\right]\label{qw}.\eeq As narrated  in the previous section, due to this spin magnetic field the spin of electron follow a Lorentz like force,  given as \beq \vec{F}(\sigma) = -\frac{q}{c}(\vec{v}\times \vec{B}^{cs}(\sigma))\label{lo},\eeq where $\vec{v}$ is the velocity of the charged particle. Due to this spin dependent Lorentz force we can achieve the transverse spin separation in the sample. This transverse spin separation is commonly known as the spin Hall effect. Similar to the classical Hall effect, due to the presence of spin dependent magnetic field, we can write the expression of fictitious defect modified spin dependent Hall electric field as
\begin{eqnarray}
\vec{E}_{H}(\sigma) &=& R_{H}(\vec{B}^{cs}(\sigma)\times\vec{j}_{c}),\label{asd}
\end{eqnarray}  where $R_{H}$ and $\vec{j}_{c}$ are the Hall coefficient and charge current respectively and can be expressed as $R_{H} = -\frac{1}{nqc},$ $\vec{j}_{c} = nq\vec{v}_{0}.$ Here $n$ and $\vec{v}_{0}$ are the concentration of the charge carriers and drift velocity respectively. Now if we deal with spin polarized electrons we can take the spin average of the spin magnetic field as
\beq \vec{B}^{'} = -\langle \vec{B}^{cs}(\sigma) \rangle = \frac{\hbar}{4mc}\left\langle\vec{\nabla}\times(\vec{\xi}\times \vec{E}_{tot})\right\rangle,\label{el}\eeq
where $\vec{\xi},$ the spin polarization vector can be written as $\vec{\xi} = \langle\vec{\sigma}\rangle,$ with absolute values lying between the range $0$ to $1.$ The spin polarization vector can also be written as
 \beq \xi = \frac{n_{+} - n_{-}}{n_{+} + n_{-}},\eeq where $n_{\pm}$ are the concentration of carriers with the spin alignment in the parallel and anti-parallel direction of the spin polarization vector. The polarization of spin is an important concept in the context of $spintronics$. The total charge concentration can be written as $n = n_{+} + n_{-}.$
 Thus with the topological defect background we can write the measurable defect modified Hall electric field as
 \begin{eqnarray}
 \vec{E}_{H} &=& R_{H}\left(\vec{B}^{'} \times \vec{j}_{c}\right)\nonumber\\
  &=& - R_{H}\left[\left(\frac{\hbar}{4mc}\left\langle\vec{\nabla}\times(\vec{\xi}\times (\vec{E}- \vec{E}_{cs}))\right \rangle\right)\times \vec{j}_{c}\right]\label{ell}.
  \end{eqnarray}

It is evident from the expressions of (\ref{el}) and (\ref{ell}) that to obtain a concrete expression of the Hall electric field  and the spin averaged magnetic field we have to calculate the volume average of $\nabla_{i}\nabla_{j}V_{l}(r).$ As $\sigma_{i}\partial_{i}\Omega_{jk}\partial_{k}V$ and $\sigma_{j}\partial_{j}\Omega_{ik}\partial_{k}V$ are very small in magnitude we have neglected the corresponding terms. We have only considered the terms having $\langle\nabla_{i}\nabla_{j}V(r)\rangle.$ Thus the defect modified Hall electric field as well as the spin average of the magnetic field consequently depend on the geometry and symmetry of the crystal. To elaborate the statement and to show how the Hall voltage depends on the system geometry we will consider a non-cubic crystal and study the effect of cosmic string parameters and non-cubic parameters on spin current.

The most symmetric crystal is the crystal having cubic symmetry. Monoclinic, orthorhombic, rhombohedral, tetragonal, hexagonal and triclinic are other less symmetric crystals, commonly known as the non-cubic crystals. A non-cubic crystal is one which have different properties in different direction.
Considering  the example of a tetragonal crystal i.e the crystal having $a = b$ with the unit vector along the crystallographic direction $c,$ ($a,~b, ~c$ are the three crystal axis) we would like to investigate the expressions of the above mentioned defect modified Hall electric field. In order to execute that, we have to first calculate the volume average \cite{n}, which for the tetragonal crystal is given by
\beq \langle \nabla_{i}\nabla_{j}V_{l}(r)\rangle = \mu\left[\kappa_{a}\delta_{ij} + (\kappa_{c} - \kappa_{a})n_{i}n_{j}\right] ,\eeq
where $\kappa_{a} \neq \kappa_{c}$ are the non-cubic parameters of order unity.
Here $\mu$ is a constant which depends on the charge and the concentration
of ions and has been calculated in \cite{n}. We know that due to the presence of the cosmic string background, the space time is deformed. In the expression of the spin averaged magnetic field, there are actually two parts i.e we can write $\vec{B}^{'} = B_{1} + B_{2},$ where $B_{1} = \frac{\hbar}{4mc}\left\langle\vec{\nabla}\times(\vec{\xi}\times \vec{E})\right\rangle$ and $B_{2} = \frac{\hbar}{4mc}\left\langle\vec{\nabla}\times(\vec{\xi}\times \vec{E}_{cs})\right\rangle,$ with $\vec{E}$ is the external electric field and $\vec{E}_{cs}$ is the defect induced electric field. Explicitly we can write,
\beq \vec{B}_{1} = \mu\frac{\hbar}{2mc}[(\kappa_{a} + \kappa_{c} )\vec{\xi} + (\kappa_{a} - \kappa_{c})(\vec{n}.\vec{\xi})\vec{n}]\eeq  and
\beq \vec{B}_{2} = \mu Tr\{\Omega\}\frac{\hbar}{4mc}[(\kappa_{a} + \kappa_{c})\vec{\xi} + (\kappa_{a} - \kappa_{c})(\vec{n}.\vec{\xi})\vec{n}],\eeq
where $Tr\{\Omega\}$ is the trace of the cosmic string matrix $\Omega,$ which can be written as
\begin{equation}\label{ome}
\Omega^{\mu}_{~~a}  = \left(\begin{array}{ccrr}
0 & 0 & 0 & 0 \\
0 & cos\varphi & sin\varphi & 0\\
0 &\frac{sin\varphi}{\eta\rho}& -\frac{cos\varphi}{\eta\rho} & 0 \\
0 & 0 & 0 & 0
\end{array} \right)
\end{equation}
By using the average of the polar angle $\varphi$ as well as the eqn (\ref{ome}), we can write the expression of the magnetic field $B_{2}$ as
\begin{eqnarray}
\vec{B}_{2} &=& \mu \frac{\hbar}{4mc}(1 - \frac{1}{\eta})[(\kappa_{a} + \kappa_{c} )\vec{\xi}
+ (\kappa_{a} - \kappa_{c})(\vec{n}.\vec{\xi})\vec{n}]\nonumber\\
&=& \mu \frac{q\zeta G}{mc}[(\kappa_{a} + \kappa_{c} )\vec{\xi} + (\kappa_{a} - \kappa_{c})(\vec{n}.\vec{\xi})\vec{n}]
,\end{eqnarray}
where $\eta = 1 - \frac{4\zeta G}{c^{2}},$ is the deficit angle and $\zeta$ is the linear mass density  of the cosmic string. In general $\eta = 1$ describe the space time without the effect of cosmic string effect. The expression of $\vec{B}_{1}$ and $\vec{B}_{2}$ clearly indicates the dependence of the spin averaged magnetic field on both the cosmic string parameters and the parameters related to the geometry of the crystal. The contribution of the cosmic string induced magnetic field $B_{2}$ on the Hall electric field is given by
\begin{eqnarray}
E_{H, cs} = R_{H}\mu \frac{\hbar\zeta G}{mc^{3}}[(\kappa_{a} + \kappa_{c} )(\vec{\xi}\times\vec{j}_{c})+ (\kappa_{a} - \kappa_{c})(\vec{n}.\vec{\xi})(\vec{n}\times\vec{j}_{c})].
\end{eqnarray}
Including the effect of $B_{1}$ and $B_{2}$ the expression for the total Hall electric field can be given by
\begin{eqnarray}\label{la}
E_{H} &=& R_{H}\mu \left(1 + \frac{2\zeta G}{c^{2}}\right)\frac{\hbar}{2mc}\big[(\kappa_{a} + \kappa_{c} )(\vec{\xi}\times\vec{j}_{c}) + (\kappa_{a} - \kappa_{c})(\vec{n}.\vec{\xi})(\vec{n}\times\vec{j}_{c})\big].
\end{eqnarray}


This is the expression of the modified Hall electric field in a non-cubic crystal in presence of the cosmic string background, which we call modified Hall field. The dependence of this Hall field on cosmic string parameters i.e on $\zeta$  and $G$  as well as on non-cubic parameters is evident
from Eqn. (\ref{la}) which  indicates that the crystal symmetry and geometry plays a crucial role in
determining such Hall fields in asymmetric semiconductors.
\section{Discussion}
In the present paper, we start with the Dirac Hamiltonian in presence of a cosmic string background and have studied the effect of cosmic string on different crutial parameters, studied in the context of spin dynamics. We build our model Hamiltonian with the help of FW transformation and derive the dependence of cosmic string parameters on SOC Hamiltonian and different phases. Due to the presence of the cosmic string space time the direction of the spin flow is banded. The presence of the cosmic string background induces some extra terms in the expression of SOC Hamiltonian and also the Zeeman term gets modified. This modification is due to the interaction of the spin of the electron with the cosmic string. But most importantly, we note that the vector potentials also contain some extra terms. These modifications of the gauge potentials basically correspond to the modified $AC$ and $AB$ phases, the interplay of which may help one to propose the spin filter configuration in our model.

Secondly, from the space part of the SU(2) gauge field, which comes from the total SOC term, the Berry curvature is derived. On the other hand, the time part of the SU(2) gauge field  gives the spin electric field. This spin dependent electric field has contribution from the background cosmic string. How the spin electric field changes in a time dependent electric field is studied which shows the dependence of the spin electric field on cosmic string parameter. As the external electric field is considered to be a oscillatory function of time, the spin electric field as well as spin current are of alternating types.

Finally, the spin Hall effect in the cosmic string background for an asymmetric crystal is also investigated. From the 
Pauli-Schrodinger equation we can derive the spin dependent magnetic field and also the expression of defect 
modified Hall voltage in our system. The Hall voltage is caused due to the polarization of the electron spin. In this 
context the importance of crystal symmetry and geometry of experiment is established in the cosmic string background. 
The spin current for non-cubic crystal can be achieved from the Hall electric field 
as spin dependent part of the velocity operator can be written as $ \vec{v}(\sigma) = \frac{q\tau}{m}E_{H}(\sigma),$
which is related to the spin current through the relation
$ j_{ik} = \frac{1}{2}nq(\sigma_{i}v_{k} + v_{k}\sigma_{i}). $ Here $\vec{\sigma}$ denotes the Pauli matrix and $\tau$ is the scattering time.

Let us choose another type of asymmetric crystal for example, a orthorhombic crystal, for which the volume average of the potential can be written in a diagonal form as
$\left\langle \nabla_{i}\nabla_{j}V_{l}(r)\right\rangle = \mu\kappa_{i}\delta_{ij},$ where $i,~j$ are the space index and $\kappa_{x}\neq~ \kappa_{y} \neq ~\kappa_{z},$ are the factors of order unity.
A straightforward calculation shows the expression of spin current in an orthorhombic crystal in presence of cosmic defect and can be written as  \beq j_{ik} = \mu\frac{\hbar\sigma_{c}}{4mc}(1 + \frac{2\zeta G}{c^{2}})\epsilon_{ikj}(\kappa_{x} + \kappa_{y} + \kappa_{z} - \kappa_{i})j_{0j}\label{12},\eeq where $\sigma_{c} = \frac{ne^{2}\tau}{m},$ is the usual charge conductivity in the Drude model approach \cite{n,bc}.
This expression of spin current depends on cosmic string parameters i.e on $\zeta$  and $G$  as well as on non-cubic parameters. This  indicates that the crystal symmetry and geometry gives an observable change in the conductivity in case of the  asymmetric semiconductor.
From (\ref{12}) one can easily note that 
the dependence of the spin conductivity on cosmic string parameters is of the order of $\frac{\zeta G}{c^{2}},$  which exactly matches with the spin conductivity mentioned in \cite{ma}. The effect of the cosmic string parameters and the symmetry related parameters of the crystal on the defect induced Hall electric field and spin current in case of a non-cubic parameters is clear from the presence of the $\eta$ ,$\zeta$ and the $\kappa_{i}$ terms. Importantly, both the cosmic string parameters as well as non-cubic parameters affect the conductivity as well as spin current of the system.

\hspace{1 cm}
\begin{center}
{\bf Acknowledgment}
\end{center}

We would like to acknowledge the anonymous referees for their valuable comments. 


\begin{thebibliography}{999}
\bibitem{bakke} K. Bakke, L. R. Ribeiro, C. Furtado and J. R. Nasci-mento, Phys. Rev. D {\bf 79}, 024008 (2009).
\bibitem{Bakke1}K. Bakke, C. Furtado and J. R. Nascimento, Eur. Phys.J. C {\bf 60}, 501 (2009); K. Bakke, C. Furtado and J. R.Nascimento, Eur. Phys. J. C, {\bf 64}, 169(E) (2009).
\bibitem{bakke2}  K. Bakke, J. R. Nascimento and C. Furtado, Phys. Rev.D,  {\bf 78}, 064012 (2008).
\bibitem{furtado} C. Furtado, B. G. C da Cunha, F. Moraes et. al., Phys. Lett. A, {\bf 195}, 90 (1994).
\bibitem{marques}G. A. Marques, C. Furtado, V. B. Bezerra and F. Moraes, J. Phys. A: Math. Gen. {\bf 34}, 5945(2001).
\bibitem{wolf}S. A. Wolf, D. D. Awschalom, R. A. Buhrman, J. M. Daughton, S. von
 Molnar, M. L. Roukes, A.Y. Chtchelkanova, and D. M.
 Treger, Science {\bf 294}, 1488 (2001).
 \bibitem{zutic}I. Zutic, J. Fabian, and S. D. Sarma, Rev. Mod. Phys.
 {\bf 76}, 323 (2004).
 \bibitem{sh1} J. E. Hirsch, Phys. Rev. Lett. {\bf 83}, 1834 (1999)(arXiv:cond-mat/9906160).
 \bibitem{matsuo} M. Matsuo , J Ieda, S Maekawa, Phys. Rev. B {\bf 87}, 115301 (2013), M. Matsuo et.al., Physical Review B  {\bf 84}, 104410 (2011).
\bibitem{cb}Debashree Chowdhury and B. Basu, Annals of Phys {\bf 329} 166--178 (2013).
\bibitem{bc}B. Basu and Debashree Chowdhury ,Annals of Phys , 335, 47, 2013.
\bibitem{bcs} B Basu, D Chowdhury, S Ghosh, Phys. Lett. A, 377, 1661, 2013.
\bibitem{sr}Debashree Chowdhury and B. Basu, Annals of Phys 339, 358 (2013).
\bibitem{perel} M. I. Dyakonov and V. I. Perel, Sov. Phys. JETP Lett. {\bf 13}: 467
(1971).
\bibitem{pro} Pratul Bandyopadhyay, Banasri Basu and Debashree Chowdhury, Proc. R. Soc. A {\bf 470}, 20130525 (2014).
 \bibitem{n}E. M. Chudnovsky, Phys. Rev. Lett.{\bf 99}, 206601 (2007).
\bibitem{noncubic}E. M. Chudnovsky,  Phys. Rev. B {\bf 80}, 153105 (2009).
\bibitem{naka}M. Nakahara, (Institute of Physics Publishing, Bristo, 1998)
\bibitem{ma} J. H. Wang, K. Ma, K. Li, Phys. Rev. A, {\bf 87}, 032107 (2013).
\bibitem{m} L.L Foldy and S.Wouthuysen, Phys Rev {\bf 78}, 29 (1950).
\bibitem{aharonov} Y Aharonov and D. Bohm, Phys. Review {\bf 115} 485 (1959).
\bibitem{rashba}E.I Rashba Sov. Phys. Solid State {\bf 2}, 1224 (1960); Y. Bychkov and E.I Rashba JETP Lett. {\bf 39}, 78 (1984).
\bibitem{hatano}N. Hatano, R. Shirasaki and H. Nakamura, Phys. Rev. A {\bf 75}, 032107 (2007).
\bibitem{casher}Y. Aharonov and A. Casher, Phys. lett. {\bf 53}, 319 (1984).
\bibitem{berry}M.V. Berry, Proc. R. Soc. London A {\bf 392}, 45 (1984).
 \bibitem{shibata} J. Shibata, H. Kohno, Phys. Rev. Lett. {\bf 102},086603, (2009) .
 \bibitem{fujita}T Fujita, M B A Jalil and S G Tan, New J of Phys {\bf 12}, 013016 (2010).

\end{thebibliography}
\end{document}